\documentclass[aps,prb,reprint,showpacs]{revtex4-2}
\usepackage{graphicx} % Include figure files
\usepackage{amsmath} % Useful mathematical tools
\usepackage{mathtools} % For underbrace
\usepackage{gensymb} % For degree symbol
\usepackage{textcomp}
\usepackage{bm} % bold mathi
\usepackage{natbib}
\usepackage{natmove}
\usepackage[colorlinks=true, linkcolor=blue, citecolor=blue, urlcolor=blue, linktoc=page, bookmarks=false, pdfstartview={FitH}, pdfborder={0 0 0.0 [3 3]}]{hyperref} % HyperLink
\usepackage{color} % For colored text
\usepackage{dcolumn} % Align table columns on decimal point
\newcolumntype{.}{D{.}{.}{2.1}}
\newcolumntype{-}{D{.}{.}{4.0}}
\usepackage{makecell}
\usepackage{multirow}
\usepackage{cleveref}
\usepackage{easyReview} % Review mode
\crefname{figure}{Fig.}{Figs}
\crefname{table}{Table}{Tables}
\crefname{equation}{Eq.}{Eqs.}
\crefname{section}{Sec.}{Secs.}
\renewcommand{\today}{\number\day \space \ifcase \month \or January\or February\or March\or April\or May\or June\or July\or August\or September\or October\or November\or December\fi \space \number\year} % Date
\begin{document}
% ========== TITLE ==========
\title{Rashba-like spin-orbit interaction and spin texture at the KTaO$_\text{3}$ (001) surface from DFT calculations}
% ====================
% ========== AUTHORS AND AFFILIATIONS ==========
\author{Vivek \surname{Kumar}}
\email[Email: ]{vivek18@iiserb.ac.in}
% --------------------
\author{Nirmal \surname{Ganguli}}
\email[Email: ]{NGanguli@iiserb.ac.in}
\affiliation{Department of Physics, Indian Institute of Science Education and Research Bhopal, Bhauri, Bhopal 462066, India}
% ====================
\date{\today}
% ========== ABSTRACT ==========
\begin{abstract}
Rashba-like spin-orbit interaction at oxide heterostructures emerges as a much sought-after feature in the context of oxide spintronics and spin-orbitronics. KTaO$_3$ (KTO) is one of the best substrates available for the purpose, owing to its strong spin-orbit interaction and alternating $+1|-1$ charged layers along the (001) direction. Employing first-principles calculations within density functional theory (DFT) and proposing a possible electrostatic model for charge transfer to the surfaces of KTO slabs, we comprehensively analyze Rashba-like spin-orbit interaction with the help of three-dimensional band dispersion, isoenergetic contours, and projected spin textures $-$ all directly obtained from our DFT results $-$ in a thin insulating slab and a conducting thick slab of KTO. Our results reveal reasonably strong linear Rashba interaction with no signature of Dresselhaus or higher order Rashba interactions in the systems considered here. The rigorous analysis presented here may be crucial for future developments in oxide spintronics.
\end{abstract}
% ====================
\maketitle

% ========== INTRODUCTION ==========
\section{\label{sec:intro}Introduction}
Oxide heterostructures have attracted remarkable attention of researchers in the last two decades owing to unprecedented advancements in synthesis techniques and the revelation of several exciting properties, including two-dimensional electron system, magnetism, superconductivity, and interesting spin-orbit entanglement physics. Besides SrTiO$_3$, KTaO$_3$ (KTO) has recently been widely used as a substrate for deposition of a perovskite oxide film due to its $+1|-1$ charged layers along $001$ direction, and strong spin-orbit interaction arising from $5d$ element Ta \cite{WadehraNC20, Garcia-CastroPRB20, ShanavasPRL14, GuptaAM22}. Spin-orbit interaction leads to a sizable Rashba-effect at a KTO slab that may help design spintronic devices in the form of a perovskite oxide heterostructure \cite{ShanavasPRL14, GuptaAM22, BaltzRMP18, ChakrabortyPRB20}. Further, a relatively large bulk band gap of 3.64~eV of KTO \cite{JellisonPRB06} can help design heterostructures with many other perovskite oxides for hosting a two-dimensional electron system at the interface with suitable surface termination, making it a popular choice for a substrate \cite{GanguliPRL14}. Experimental studies of a few KTO-based heterostructures with an interfacial two-dimensional electron system revealed some fascinating physical properties \cite{Zou2015a, WadehraNC20, Goyal2020, Zhang2017, Zhang2018, Kumar2021}. \citet{WadehraNC20} reported planner Hall effect and anisotropic magnetoresistance in LaVO$_3|$KTaO$_3$ heterostructure assuming the presence of Rashba-like spin-orbit interaction. The two-dimensional conducting layer at the interface of (001)-oriented EuO$|$KTO shows a high degree of spin polarization manifested as negative magnetoresistance, quantum oscillations, and anomalous Hall effect \cite{Kumar2021}. A heterostructure of LaCrO$_{3}$ and KTO exhibits a two-dimensional conducting system at the interface, showing interfacial coupling \cite{Al-Tawhid2021}. Recently two-dimensional electronic properties of magnetic oxyfluoride superlattice on KTO substrate were reported from theoretical calculations \cite{Garcia-CastroPRB20}. Additionally, spin-polarized electron transport, persistent photocurrent, topological Hall effect, and inverse Edelstein effect may be realized in heterostructures involving KTO \cite{GuptaAM22}. The above discussion highlights the importance of studying Rashba-like spin-orbit interaction in the substrate material KTO.

A few experimental and theoretical investigations attempted to understand the exciting features of KTO and its importance as a substrate. \citet{SetvinS18} reported a rearrangement of the top layer of a KTO slab into KO and TaO$_2$ stripes to compensate for a polar catastrophe. Angle-resolved photoemission spectroscopy studies found two-dimensional electron gas (2DEG) at the (100) and (001) surfaces of KTO $-$ a relatively large gap band insulator \cite{King2012, Santander-Syro2012} and Rashba-like spin-orbit interaction at the Fermi level \cite{King2012}. The polar nature of the KTO slab may explain the origin of the conducting system, similar to that of LaAlO$_3|$SrTiO$_3$ interface \cite{GanguliPRL14}. Alternatively, an uncompensated charge on the KTO surface may lead to complex electronic states like charge density waves coexisting with strongly-localized electron polarons and bipolarons \cite{ReticcioliNC22}. Superconductivity may reportedly be induced on the surface of an insulating crystal of KTO \cite{Ueno2011}. Oxygen-vacant conducting KTO crystals via Ar$^+$ irradiation revealed quantum oscillations in magnetoresistance \cite{Harashima2013}. A large $k^3$-Rashba type splitting at the oxyfluoride interfaces involving KTO was reported and asserted to be larger than that of LaAlO$_3|$SrTiO$_3$ interface \cite{Garcia-CastroPRB20}. Currently, the community is unsure of the possibility of a $k^3$ Rashba-like interaction at KTO slab \cite{King2012, KimPRB14}. A few research groups theoretically investigated an insulator to metal transition and spin-orbit interaction driven Rashba effect at 001 surfaces of KTaO$_3$ \cite{ShanavasPRL14, Wu2020}. However, several issues remain unaddressed, such as (a) if the KTO slab becomes conducting only beyond a critical thickness, (b) if true, any estimate of such a thickness, and (c) if Dresselhaus or $k^3$-Rashba interaction may be relevant for the system. Moreover, no confirmatory characterization for the nature of Rashba-like interaction has been reported for KTO systems, although this is essential from an application point of view.

In order to address the issues mentioned above, we use {\em ab initio} density functional calculations in combination with analytical and numerical modeling to thoroughly investigate the physical properties of KTO in its bulk form as well as thin and thick slabs by examining a possible conducting state and spin-orbit-driven properties. We developed a theoretical model based on electrostatics to assess if a KTO slab would be conducting only beyond a critical thickness. Besides studying the band structure to analyze Rashba-like physics, we take a closer look into the relevant 3D bands, isoenergetic contours, and spin textures directly obtained from our density functional theory (DFT) calculations offering a conclusive understanding of the highly sought-after substrate in oxide electronics. Our work is presented in the following order: \cref{sec:method} details the structural models considered for the calculations and the methods and techniques used for the calculations. Our results are discussed in \cref{sec:results}. Finally, we summarize the important findings and conclude the article in \cref{sec:conc}.

% ========== METHOD ==========
\section{\label{sec:method}Structure and method of calculations}
In its bulk form, KTaO$_3$ crystallizes in an inversion symmetric space group $Pm\bar{3}m$ having a cubic structure at room temperature. Six O atoms surround each Ta atom, forming an octahedron. Starting from bulk KTO for reference, we have simulated two unit cell thick and twenty unit cell thick KTO slabs to study the implications of Rashba-like spin-orbit interaction on the insulating and conducting surfaces in detail.

All the total energy and electronic structure calculations presented here are carried out using density functional theory as implemented in the {\scshape vasp} code \cite{vasp1,vasp2}. The projector augmented wave (PAW) method \cite{paw} is employed for describing the potential, in conjunction with a plane wave basis set with 500~eV energy cutoff for expanding the wavefunctions. The exchange-correlation functional is treated within local (spin) density approximation (L(S)DA) \cite{ldaCA,PerdewPRB81} along with Hubbard-$U$ correction for electron-electron correlation in Ta-$5d$ states with $U_\text{eff} = U - J = 1$~eV; the so-called LDA$+U$ method \cite{DudarevPRB98}. The modest $U$-value for describing the Ta-$5d$ states avoids overcorrection as the Ta-$5d$ states are known to be not so strongly correlated and host no magnetism. Our tests with higher $U$-values reveal a qualitatively similar electronic structure. Introducing a vacuum of at least 20~\AA\ along the (001) direction to sufficiently separate the periodic images of the slabs, the integration over the Brillouin zone is performed using a $19 \times 19 \times 1$ $\Gamma$-centered $k$-mesh within the improved tetrahedron method \cite{BlochlPRB94T}. The atomic positions are optimized by minimizing the Hellman-Feynman force on each atom to a tolerance of $10^{-2}$~eV~\AA$^{-1}$. A small electronic convergence threshold of $10^{-7}$~eV is used for spin-orbit interaction calculations.

% ========== RESULTS AND DISCUSSIONS ==========
\section{\label{sec:results}Results and discussions}
In its bulk form, KTaO$_3$ is found in a nominal oxidation state K$^+$Ta$^{5+}$O$^{2-}_3$. While the bulk material shows a band gap of 3.64~eV \cite{JellisonPRB06}, the surfaces of a thick slab of KTO become conducting, owing to the polar arrangement of alternating (K$^+$O$^{2-}$)$^-$ and (Ta$^{5+}$O$^{2-}_2$)$^+$ layers along (001)-direction \cite{Santander-Syro2012, GanguliPRL14, ShanavasPRL14}. Therefore, to understand the essential features of KTO as a substrate, we simulate and critically analyze the properties of (a) bulk KTO, (b) a thin slab of KTO with 2 unit cell (uc) thickness, and (c) a thick slab of KTO with 20 uc thickness. The results of our systematic calculations are discussed here.

\subsection{\label{sec:bulkKTO}Bulk KTO}
% ========== Figure: Bulk KTO DoS band ==========
\begin{figure}
	\includegraphics[scale=0.35]{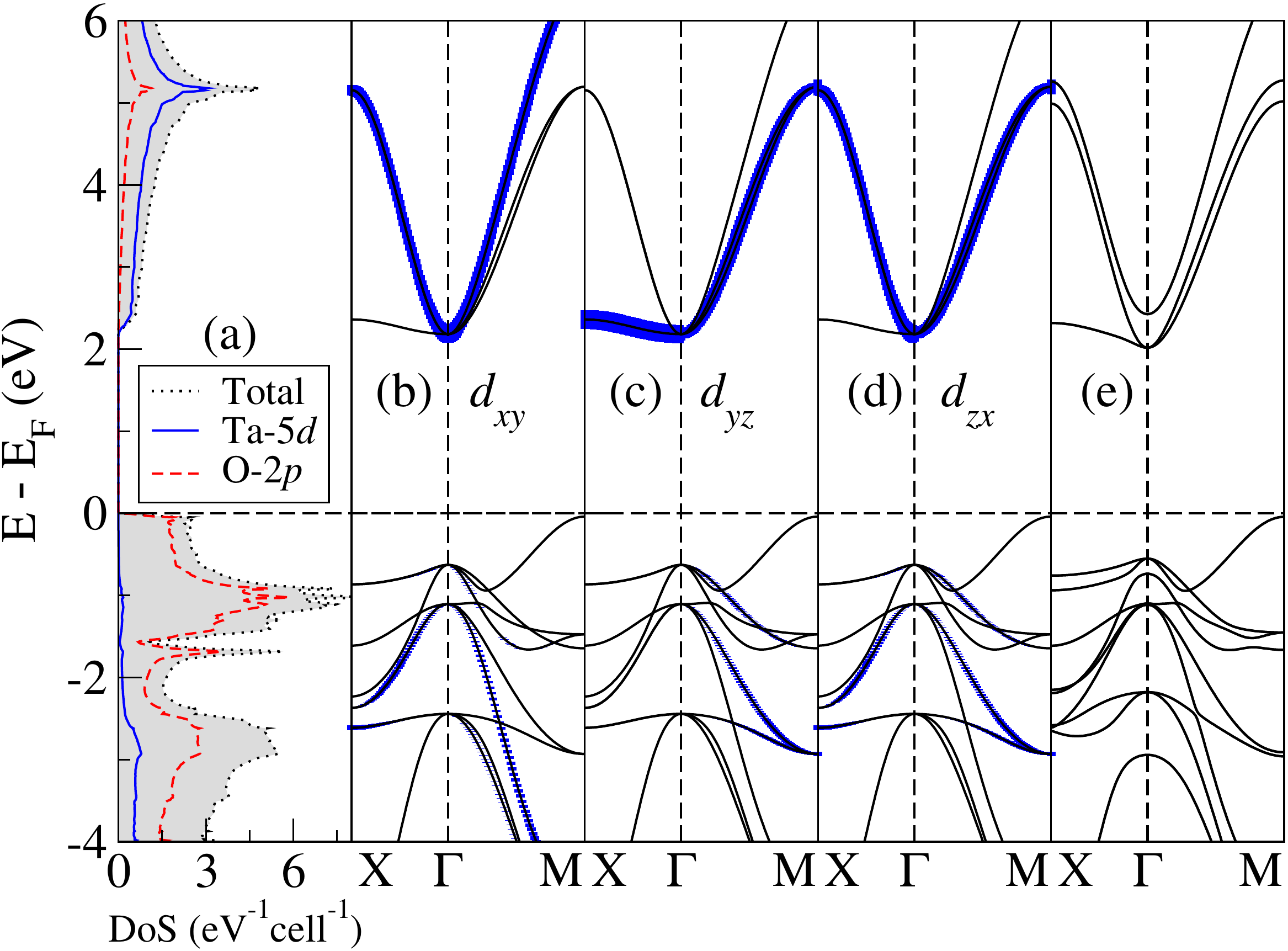}
	\caption{\label{Fig:KTOBulk-OrbitalsandSoc}Spin-unpolarized density of states (DoS) along with projected DoS for Ta-$5d$ and O-$2p$ orbitals for bulk KTO is shown in panel (a). The spin-unpolarized band structure of bulk KTO along $X \to \Gamma \to M$ direction with the dominant contributions of Ta-$d_{xy}$, $d_{yz}$, and $d_{zx}$ orbitals are highlighted and displayed in (b), (c), and (d), respectively. Panel (e) shows the band dispersion upon considering spin-orbit interaction in our calculation.}
\end{figure}
% ========== ==========
Although Ta-$5d$ states remain empty in bulk KTO, the rich physics in KTO as a substrate or slab arises from the partially filled Ta-$5d$ states, making them worth investigating. Subject to an octahedral crystal field from the surrounding oxygen atoms, Ta-$5d$ states split into a lower-energy threefold degenerate $t_{2g}$ manifold and a higher energy twofold degenerate $e_{g}$ manifold. The spin-unpolarized density of states (DoS) for bulk KTO shown in \cref{Fig:KTOBulk-OrbitalsandSoc}(a) captures a sizable band gap of more than 2~eV within DFT calculation, with Ta-$5d$ states above Fermi level, having some admixture with O-$2p$ states, as seen from the projected DoS of Ta-$5d$ and O-$2p$ orbitals. The orbital characters of $d_{xy}$, $d_{yz}$, and $d_{zx}$ highlighted in \cref{Fig:KTOBulk-OrbitalsandSoc}(b),(c),(d), respectively, suggests an intuitive dispersion of the bands along the high-symmetry $k$-path along $X \to \Gamma \to M$, with an overall degeneracy in the relevant energy range. Upon considering spin-orbit interaction, we notice that although multiple bands below and above the Fermi level have degeneracy lifted in a certain range of $k$-points (see \cref{Fig:KTOBulk-OrbitalsandSoc}(e)), no Rashba-like splitting is observed due to inversion symmetry being preserved. However, inversion symmetry is compromised in a slab geometry, along with the development of an electric field due to the polar nature of KTO along (001) direction, which may lead to Rashba-like interaction.

\subsection{\label{sec:slab}KTO slab}
% Charge transfer
Since a KTO slab arranges itself in $+1|-1$ charged layers along (001) direction, an electrostatic potential build-up leading to charge transfer may be realized \cite{GanguliPRL14}. We develop a simple parallel plate capacitor model similar to that in ref~\cite{GanguliPRL14} to understand the origin of the experimentally observed conducting surfaces in the KTO slab.
% ========== Figure: Schematic potential ==========
\begin{figure}
    \centering
    \includegraphics[scale = 0.17]{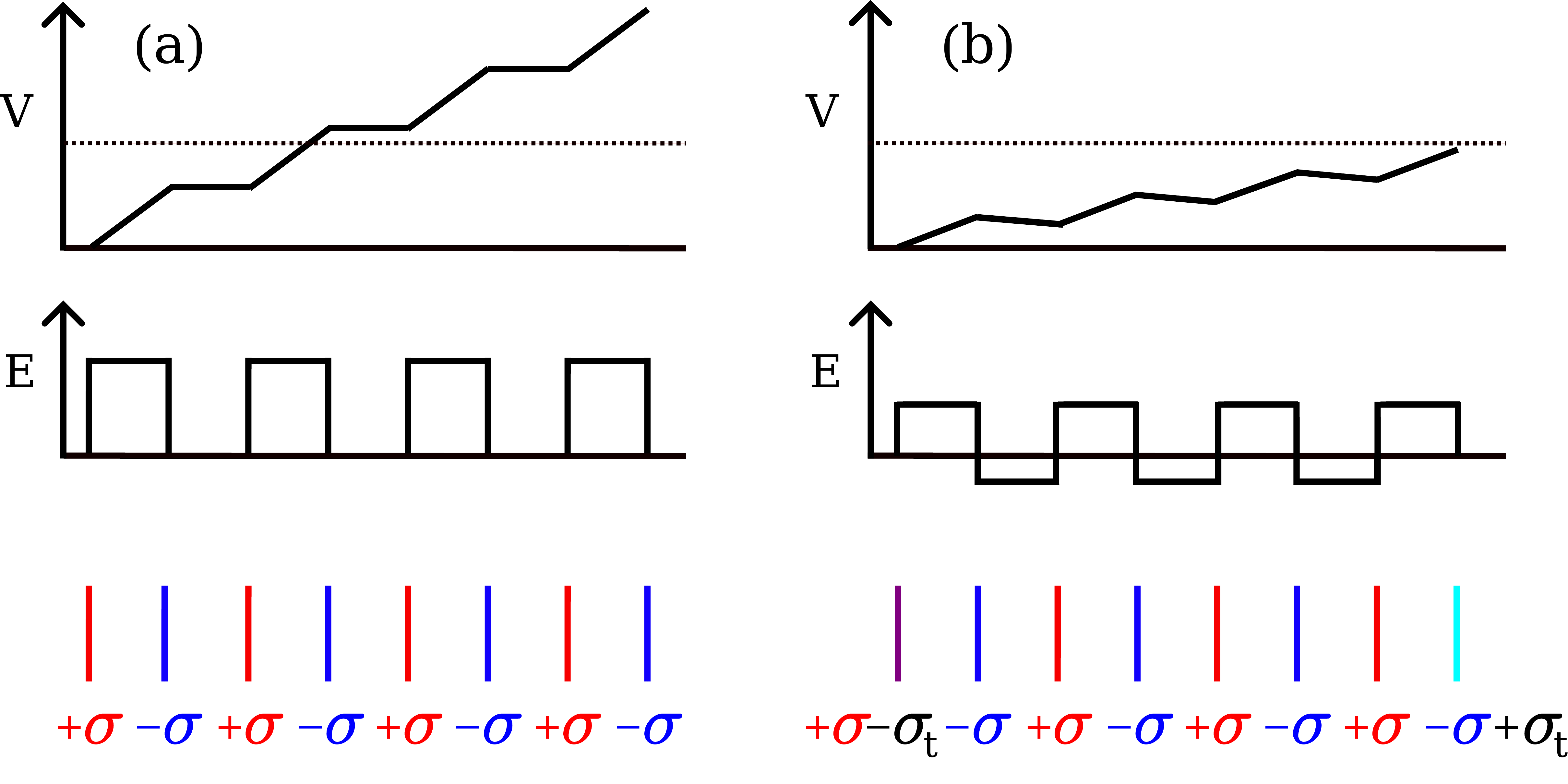}
    \caption{\label{fig:SchematicPotential}Panel (a) schematically illustrates the planes with $\pm \sigma$ surface charge density (bottom), the corresponding electric field (middle), and a monotonically increasing potential (top) before any charge reconstruction. Panel (b) depicts the electric field and electrostatic potential after a charge transfer of density $\pm \sigma_t$. The dotted horizontal line represents the band gap.}
\end{figure}
% ========== ==========
% ========== Figure: Thin slab DoS band ==========
\begin{figure}
	\includegraphics[scale=0.35]{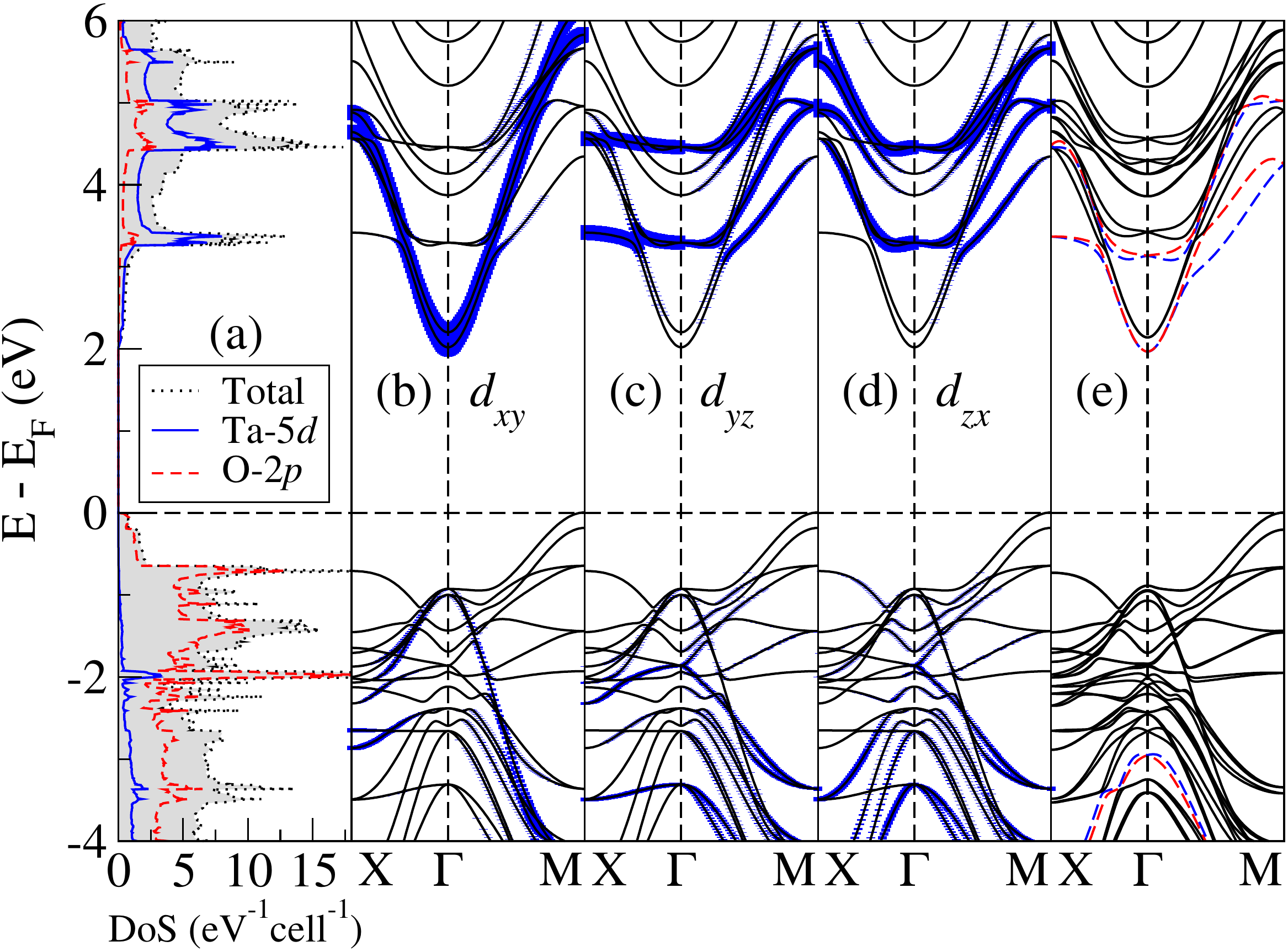}
	\caption{\label{Fig:KTOslab-OrbitalsandSoc}Spin-unpolarized DoS along with projected DoS for Ta-$5d$ and O-$2p$ orbitals for thin KTO slab is shown in panel (a). The spin-unpolarized band structure of thin KTO slab along $X \to \Gamma \to M$ direction with the dominant contributions of Ta-$d_{xy}$, $d_{yz}$, and $d_{zx}$ orbitals are highlighted and displayed in (b), (c), and (d), respectively. Panel (e) shows the band dispersion upon considering spin-orbit interaction in our calculation.}
\end{figure}
% ========== ==========
% ========== Figure: 3D bands and Rashba; Thin KTO ==========
\begin{figure*}
	\includegraphics[scale=0.64]{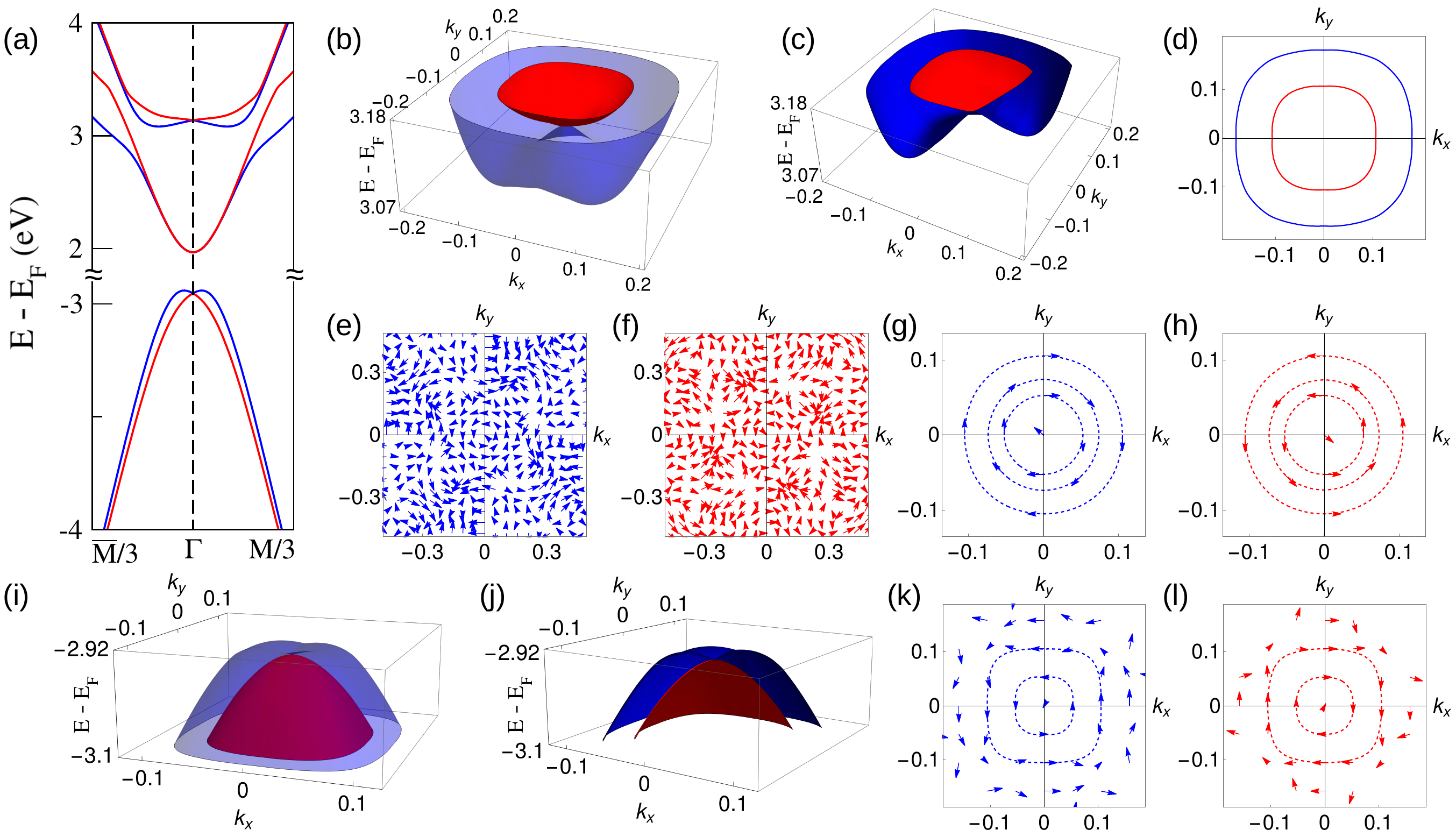}
	\caption{\label{Fig:insulatorKTO1}The bands with Rashba-like splittings for thin KTO slab, shown in (a), are closely analyzed here. The energy dispersion with $k_x$ and $k_y$, the so-called 3D band structure, for the top pair of bands is displayed in (b) and a cross-section of the same is shown in (c). Panel (d) shows the isoenergetic contours corresponding to the same pair of bands with $E - E_F = 3.18$~eV. Panels (e) and (f) depict projected spin vectors $\langle S_x \rangle \hat{x} + \langle S_y \rangle \hat{y}$ in $k_x$-$k_y$ plane, as obtained from our DFT calculations, in a relevant energy range for the bands, revealing helical spin arrangements of opposite direction in (e) and (f). The pair of bands at the middle of panel (a) do not clearly show a Rashba-like splitting, however, the isoenergetic contours and the corresponding projected spin vectors shown in (g) and (h) clearly suggest a linear Rashba interaction for the bands. The 3D bands and cross-section in (i) and (j), respectively, and the isoenergetic contours with projected spin vectors in (k) and (l) correspond to the Rashba-like split bands at the bottom of the panel (a).}
\end{figure*}
% ========== ==========
A system of $n$ number of KTO unit cells along (001) direction in a slab with alternating $+1|-1$ charged layers may be represented as an arrangement of parallel plates with $+\sigma|-\sigma$ uniform surface charge densities for the present purpose. \cref{fig:SchematicPotential}(a) schematically depicts the arrangement of parallel plates with alternating $+\sigma|-\sigma$ surface charge density, the corresponding electric field, and a monotonic development of electrostatic potential from bottom to top. For $n$ number of unit cells along (001) direction, the potential build-up may be given by $V = na\sigma/2\epsilon_\text{KTO}$, where $a$ and $\epsilon_\text{KTO}$ represent the lattice constant and permittivity of KTO, respectively. For a large number of KTO unit cells, the electrostatic potential development would surpass the band gap potential of KTO (notionally illustrated in \cref{fig:SchematicPotential} with a horizontal dotted line), compelling the system to reorganize the charge distribution. Transferring some opposite charge to both surfaces, thus limiting the electrostatic potential development below the band gap potential, could be a plausible solution. Using the principle of superposition in electrostatics, one can express this plausible solution as
\begin{equation}
    \varepsilon_g \geq \frac{na\sigma}{2\epsilon_\text{KTO}} - \frac{na\sigma_t}{\epsilon_\text{KTO}} = \frac{na(\sigma - 2\sigma_t)}{2\epsilon_\text{KTO}}, \label{eq:ChargeTransfer}
\end{equation}
where $\varepsilon_g$ represents the band gap potential and $\sigma_t$ corresponds to the transferred charge density on the surfaces, as illustrated in \cref{fig:SchematicPotential}(b). \cref{eq:ChargeTransfer} indicates that the potential development will be completely arrested for $\sigma_t = \sigma/2$, a requirement for a KTO slab of effectively infinite thickness. Assuming appropriate values for $\epsilon_\text{KTO}$ \footnote{While the dielectric constant of KTO varies over a large range with temperature, we have considered a reasonable value of 220 \cite{AgrawalJPCSSP70}.} and $\sigma$, a charge transfer to the surfaces of the slab would be warranted for $n \geq 20$ considering the band gap of KTO obtained from our calculations. In contrast, $n \geq 36$ may be required to warrant a charge transfer considering the experimental band gap. Although the dielectric constant of KTO may differ depending on the experimental conditions that may alter our predictions, we could not find any experimental report on the thickness-dependent conducting nature of the KTO slab, keeping us from verifying our predictions. We note that while offering a possible physical explanation for the conducting nature of a thick KTO slab, our model may not be the only possible mechanism that explains the conducting nature of a thick KTO slab. To understand the physical properties of insulating and conducting KTO slabs at optimum computational cost, we have considered a thin (2 uc thick) and a thick (20~uc thick) KTO slab for further calculations.
 
\subsubsection{\label{sec:ThinKTO}Thin KTO slab}
We first consider a thin KTO slab of 2~uc thickness that, as discussed above, is expected to behave as an insulator. This system's density of states and band dispersion are shown in \cref{Fig:KTOslab-OrbitalsandSoc}. We gather from the spin-unpolarized DoS shown in \cref{Fig:KTOslab-OrbitalsandSoc}(a) that compared to bulk KTO (see \cref{Fig:KTOBulk-OrbitalsandSoc}(a)), the band gap marginally reduces, with tails near the valence band maximum and the conduction band minimum. The projected DoS suggests that while the bands above the Fermi level exhibit a strong Ta-$5d$ character hybridized with O-$2p$ character, the ones below the Fermi level are mainly derived from O-$2p$ states with a bit of admixture of Ta-$5d$ states in the lower energy part. Further, a comparison of \cref{Fig:KTOslab-OrbitalsandSoc}(b),(c),(d) with \cref{Fig:KTOBulk-OrbitalsandSoc}(b),(c),(d) reveals that for the slab the Ta-$5d_{yz}$ and Ta-$5d_{zx}$ bands have moved higher in energy due to exposed (001) surfaces and lack of periodicity along the $c$-direction, making the conduction band minimum primarily of Ta-$5d_{xy}$ character. Upon considering spin-orbit interaction in our calculations, we notice Rashba-like splitting in some of the bands, as highlighted in \cref{Fig:KTOslab-OrbitalsandSoc}(e), warranting a detailed investigation of Rashba-like spin-orbit interaction in the system.

We closely analyze the implications of spin-orbit interaction on the highlighted bands by exhibiting only them in \cref{Fig:insulatorKTO1}(a), omitting the other bands. The top pair of bands having predominantly Ta-$5d_{yz}/5d_{zx}$ projections (see \cref{Fig:KTOslab-OrbitalsandSoc}(c),(d)) show a clear split in the momentum space along $\bar{M} \to \Gamma \to M$ direction, according to the convention in ref \cite{ChakrabortyPRB20}. The energy dispersion for these bands as a function of $k_x$ and $k_y$ in the entire Brillouin zone (the so-called 3D bands) and a cross-section of the 3D bands with $k_x = k_y$ plane are displayed in \cref{Fig:insulatorKTO1}(b) and \cref{Fig:insulatorKTO1}(c), respectively. Isoenergetic contours of these bands for $E - E_F = 3.18$~eV displayed in \cref{Fig:insulatorKTO1}(d) are well-separated from each other but do not have a circular shape, understandably due to the nature of dispersion of $d_{yz}$ and $d_{zx}$ bands in the $k_x$-$k_y$ plane. To examine the possibility of any helical spin arrangement for the bands, we extracted from our DFT results the projected spin components $\langle S_x \rangle$ and $\langle S_y \rangle$, and plotted the projected spin vectors $\langle S_x \rangle \hat{x} + \langle S_y \rangle \hat{y}$ in the $k_x$-$k_y$ plane over a relevant energy range, as displayed in \cref{Fig:insulatorKTO1}(e) and \cref{Fig:insulatorKTO1}(f). The figures reveal the highly helical arrangement of spins forming a nice texture, indicating a Rashba-like interaction in the system. As DFT results often mix up bands, concluding the exact nature of the Rashba-like interaction becomes difficult for this pair of bands. Next, we critically examine the pair of bands at the middle part of \cref{Fig:insulatorKTO1}(a). The orbital characters highlighted at \cref{Fig:KTOslab-OrbitalsandSoc}(b) indicate these bands to have predominantly Ta-$5d_{xy}$ projections. Although hardly any Rashba-like splitting is visible for this pair of bands, the corresponding circular isoenergetic contours for various energies along with the projected spin vectors tangential to the contours displayed in \cref{Fig:insulatorKTO1}(g),(h) suggest a linear Rashba-type spin-orbit interaction. Finally, the bottom pair of bands in \cref{Fig:insulatorKTO1}(a) having dominant O-$2p$ character and little admixture with Ta-$5d$ character also show a prominent Rashba-like splitting. The corresponding 3D bands and cross-section of the 3D bands are displayed in \cref{Fig:insulatorKTO1}(i) and \cref{Fig:insulatorKTO1}(j), respectively, while isoenergetic contours for several energy values along with the projected spin textures are depicted in \cref{Fig:insulatorKTO1}(k),(l). We observe a clear separation of the bands in momentum space and helical spins aligned almost tangentially to the nearly square-shaped isoenergetic contours, suggesting linear Rashba-type interaction.

After analyzing Rashba-like interactions in a thin KTO slab in sufficient detail, we consider a thick KTO slab to understand its physical properties related to spin-orbit interaction.

\subsubsection{\label{sec:ThickKTO}Thick KTO slab}
% ========== Figure: DoS, Bands for thick KTO ==========
\begin{figure}
	\includegraphics[scale=0.35]{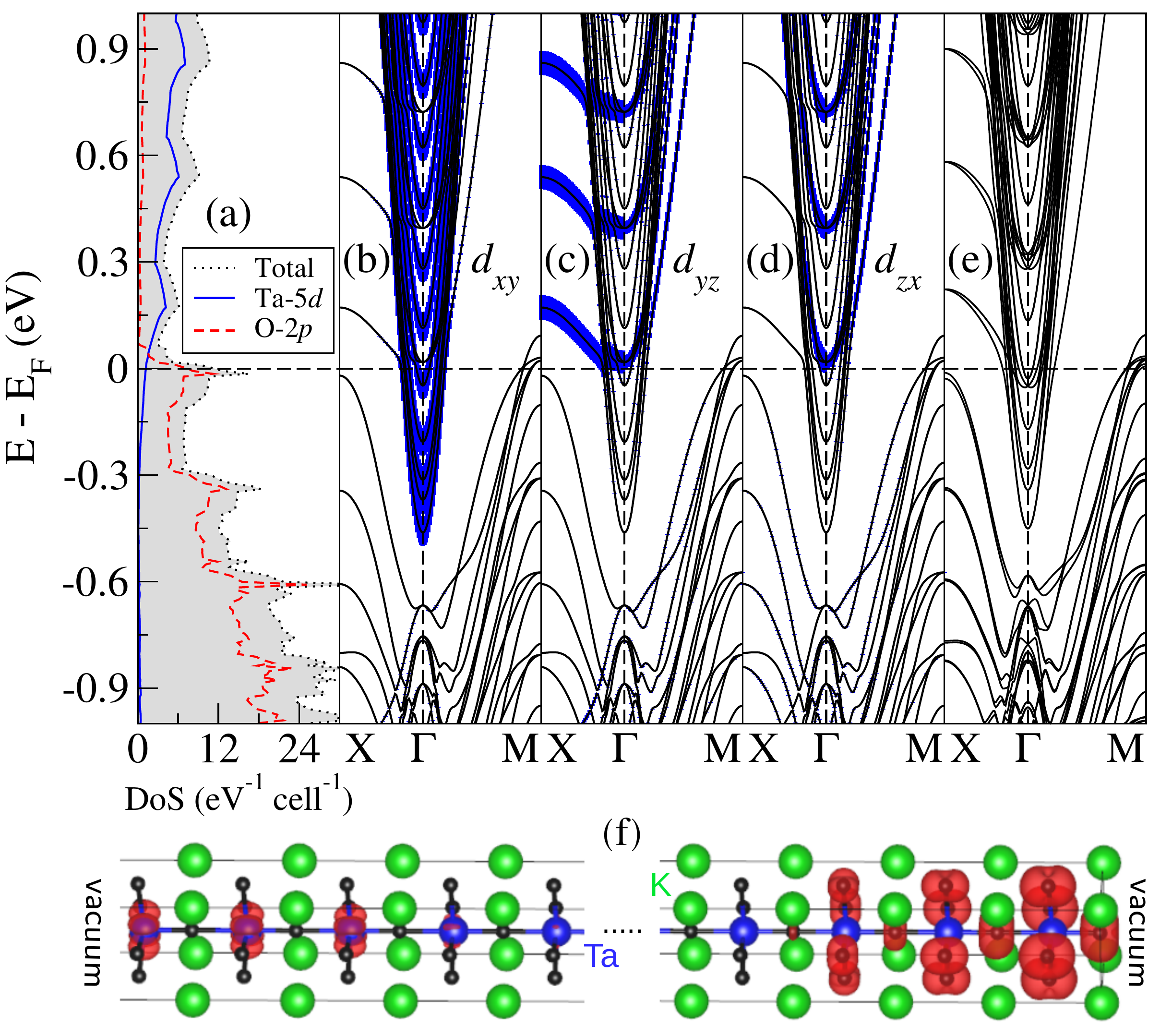}
	\caption{\label{Fig:fatbandsurf20}Spin-unpolarized DoS along with projected DoS for the 20~uc thick KTO slab is shown in panel (a). The spin-unpolarized band structure of thick KTO slab along $X \to \Gamma \to M$ direction with the dominant contributions of Ta-$d_{xy}$, $d_{yz}$, and $d_{zx}$ orbitals are highlighted and displayed in (b), (c), and (d), respectively. Panel (e) shows the band dispersion upon considering spin-orbit interaction in our calculation. The projected charge density for an energy range $[-0.1, 0.0]$~eV relative to the Fermi level (conducting electron density) is depicted in (f), omitting the featureless middle part of the simulation cell. The structural model in panel (f) is prepared with {\scshape vesta} software \cite{VESTA}.}
\end{figure}
% ========== ==========
% ========== Figure: 3D bands and Rashba; Thick KTO ==========
\begin{figure*}
	\includegraphics[scale=0.64]{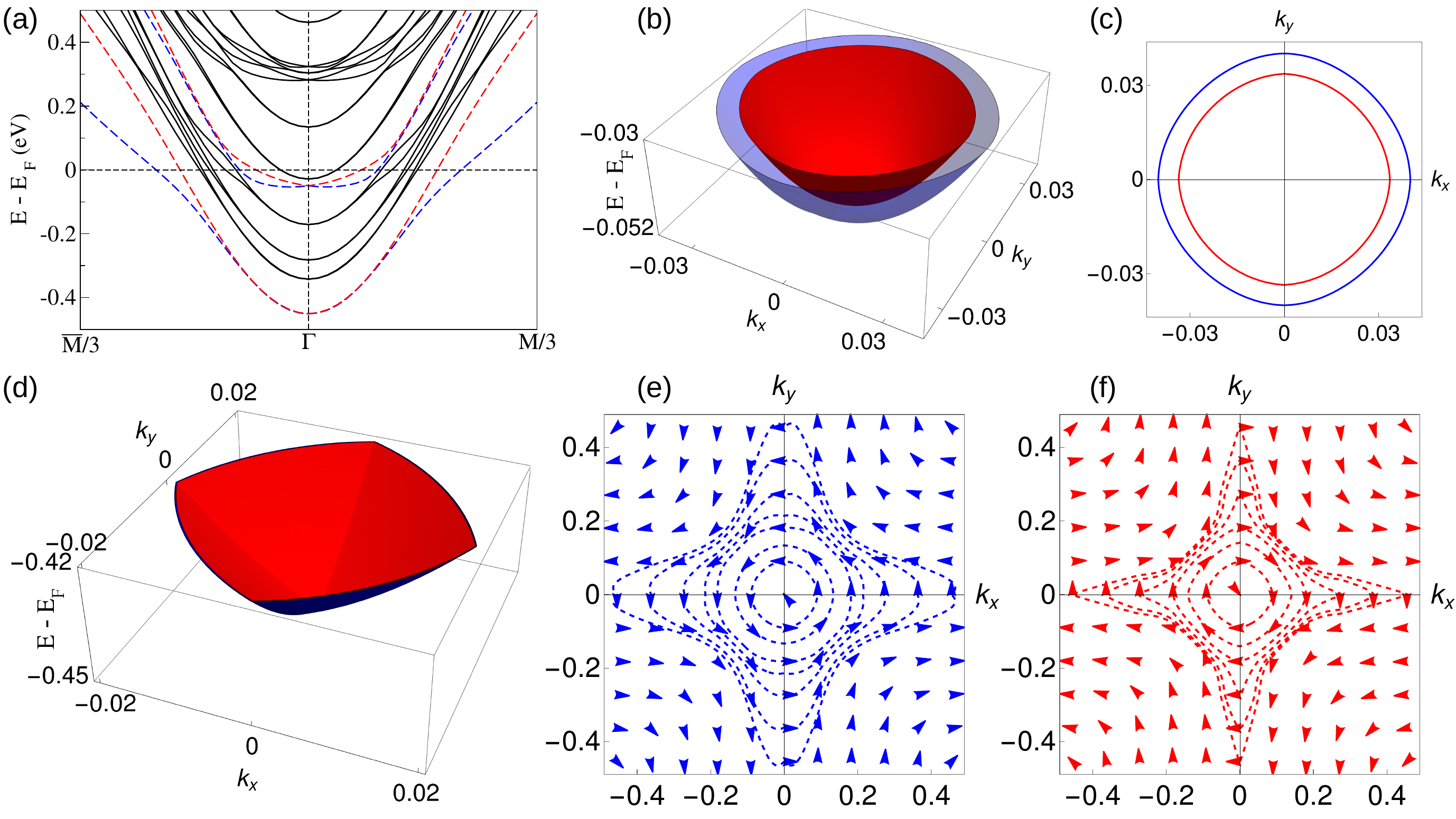}
	\caption{\label{Fig:Spins-and-band-conKTO}The conduction bands for 20~uc thick KTO slab along $\bar{M}/3 \to \Gamma \to M/3$ line segment is displayed in (a), with two pairs of Rashba-like bands highlighted. The energy dispersion as a function of $k_x$ and $k_y$ for the upper pair of highlighted bands in (a) and isoenergetic contours at $E - E_F = -0.03$~eV for the same pair of bands are shown in (b) and (c), respectively. Panel (d) exhibits the dispersion of the lower pair of highlighted bands in (a) as a function of $k_x, k_y$. Isoenergetic contours for different energies along with the projected spin vectors obtained from our DFT calculations for the same pair of bands are given in (e) and (f).}
\end{figure*}
% ========== ==========
As discussed earlier, we expect a sufficiently thick KTO slab to behave as a conductor so that a monotonic development of electrostatic potential may be averted. The results of our DFT calculations with a 20~uc thick slab indeed reveal a conducting nature of the bands, as observed from the spin-unpolarized DoS in \cref{Fig:fatbandsurf20}(a). We gather from the projected DoS that the bands below the Fermi level have a major contribution from O-$2p$ orbitals. In contrast, those above the Fermi level have a major contribution from Ta-$5d$ orbitals. The spin-unpolarized band dispersion along $X \to \Gamma \to M$ with the contributions from Ta-$5d_{xy}$, Ta-$5d_{yz}$, and Ta-$5d_{zx}$ orbitals are highlighted and displayed in \cref{Fig:fatbandsurf20}(b), \cref{Fig:fatbandsurf20}(c), and \cref{Fig:fatbandsurf20}(d), respectively. Upon considering spin-orbit interaction in our calculations, we find broken degeneracy at certain bands and Rashba-like splitting in some bands crossing the Fermi level. Before critically analyzing the Rashba-like interaction in the system, we attempt to spatially locate the conducting electrons in the slab in order to verify our assumption in the electrostatic model that some charge gets transferred to the surfaces to avoid a monotonic potential development, leading to a conducting system. A projected charge density plot for an energy range $[-0.1, 0.0]$~eV relative to the Fermi level displayed in \cref{Fig:fatbandsurf20}(f) reveals the conduction electrons to be located near the surfaces of the slab, spilling a bit on the nearby layers, and having no trace near the middle of the slab. Further, we observe the charge density around Ta atoms near the TaO$_2$-terminated surface and O atoms near the KO-terminated surface, suggesting some electron transfer towards the positively charged surface with Ta$^{5+}$O$_2^{2-}$-termination, in agreement with our model diagram shown in \cref{fig:SchematicPotential}. Partly unoccupied O-$2p$ bands near the $M$-point of the Brillouin zone testify to positive charge transfer towards the KO-terminated surface.

Subsequently, we carefully analyze Rashba-like spin-orbit interaction in the conduction bands of the 20~uc thick KTO slab. \cref{Fig:Spins-and-band-conKTO}(a) shows the conduction bands of the system along $\bar{M}/3 \to \Gamma \to M/3$ line segment, highlighting two pairs of Rashba-like bands. Comparing with \cref{Fig:fatbandsurf20}(a),(b),(c),(d), we gather that while the upper pair of bands are primarily derived from Ta-$5d_{yz}$ and Ta-$5d_{zx}$ orbitals, the lower pair of bands may predominantly correspond to hybridized Ta-$5d_{xy}$ and O-$2p$ orbitals. \cref{Fig:Spins-and-band-conKTO}(b) displays the 3D bands corresponding to the upper pair of highlighted bands, revealing a large separation between the bands everywhere in the momentum space. The corresponding isoenergetic contours for $E - E_F = -0.03$~eV, shown in \cref{Fig:Spins-and-band-conKTO}(c), indicates almost circular geometry. Moving onto the lower pair of highlighted bands in \cref{Fig:Spins-and-band-conKTO}(a), we have shown the corresponding 3D band dispersion in \cref{Fig:Spins-and-band-conKTO}(d), indicating a non-elliptical dispersion. Isoenergetic contours plotted for various energies along with the projected spin texture for the lower pair of bands, shown in \cref{Fig:Spins-and-band-conKTO}(e) and \cref{Fig:Spins-and-band-conKTO}(f), suggest that the contours for the outer band change its geometry from circular to squarish to hyperbolic with increasing energy (see \cref{Fig:Spins-and-band-conKTO}(e)), while that of the inner band transforms from squarish to hyperbolic with no circular geometry even to the lowest relevant energy. The geometry of the 3D bands (see \cref{Fig:Spins-and-band-conKTO}(d)) and the contours (see \cref{Fig:Spins-and-band-conKTO}(e) and \cref{Fig:Spins-and-band-conKTO}(f)) manifest the effect of hybridization of Ta-$5d$ and O-$2p$ orbitals. However, the projected spin vectors in \cref{Fig:Spins-and-band-conKTO}(e) and \cref{Fig:Spins-and-band-conKTO}(f) always align tangentially to the contours, in line with linear Rashba spin texture. Besides the highlighted ones, we found the other bands visible in \cref{Fig:Spins-and-band-conKTO}(a) exhibiting linear Rashba-like spin texture. Our results do not point to any cubic Rashba or Dresselhaus-like spin-orbit interaction in the KTO slabs simulated here. We note that \citet{ZhongPRB13} argued in the context of LaAlO$_3|$SrTiO$_3$ heterostructure that the multi-orbital nature of complex oxides leads to a cubic Rashba interaction. However, \citet{KimPRB14} suggested the importance of $C_{4v}$ symmetry for cubic Rashba interaction in perovskite oxides. We estimate Goldschmidt tolerance factor $t$ for packing a KTO crystal as
\begin{equation}
	t = \frac{r_{\text{K}^+} + r_{\text{O}^{2-}}}{\sqrt{2}(r_{\text{Ta}^{5+}} + r_{\text{O}^{2-}})} = 0.964,
\end{equation}
$r$ representing the ionic radii. We expect no structural distortion arising due to packing for $0.9 < t < 1$. While tetragonal distortions are common in SrTiO$_3$ and octahedral tilts may also be observed upon forming heterojunctions \cite{GanguliPRL14, ChakrabortyPRB20}, the cubic structure of KTO considered here hosts no such distortion; hence, possesses a higher symmetry than $C_{4v}$, leading to no cubic Rashba interaction.

A comparison between the spin textures of the thin and thick KTO slabs for similar bands indicates no qualitative difference near the center of the Brillouin zone, although they somewhat differ when approaching the zone boundary, owing to the mixing of bands.

% ========== CONCLUSION ==========
\section{\label{sec:conc}Conclusion}
To conclude, we use {\em ab initio} calculations within the framework of density functional theory combined with analytical and numerical modeling to critically investigate Rashba-like physics in bulk KTaO$_3$ as well as thin and thick slabs of the same material. Since the bulk structure of KTO preserves inversion symmetry, our calculations reveal no Rashba-like splitting of the bulk bands, as expected. Upon considering a slab of KTO, the alternate $+1|-1$ charged planes along (001) direction lead to an electric field and a monotonically increasing electrostatic potential. We discussed a possible electrostatic model requiring charge transfer at the surfaces beyond a critical slab thickness to avert such a monotonic potential development. Our DFT results corroborate our model, revealing an insulating slab for a small thickness of 2~uc and a conducting one with conduction electrons located near the surfaces for a relatively large thickness of 20~uc. Our results for both thin and thick KTO slabs indicate clear signatures of Rashba-like interaction. Further analysis of the Rashba-like bands shows unconventional shapes of the isoenergetic contours that may be attributed to the hybridization of different orbital characters and the dispersion of such bands in the $k_x$-$k_y$ plane. Considering the projected spin vectors obtained from our DFT calculations align tangentially to the isoenergetic contours, we interpret the underlying spin-orbit interaction to be linear Rashba type, with no clear indication of Dresselhaus or higher order Rashba interactions. The comprehensive understanding of Rashba-like spin-orbit interaction obtained via rigorous analysis of 3D bands, isoenergetic contours, and projected spin textures in one of the most promising substrates {\em viz.} KTO in different geometries presented here may unfold to be crucial in designing and characterizing oxide heterostructures for spintronic applications.

% ========== ACKNOWLEDGMENT ==========
\begin{acknowledgments}
We acknowledge financial support from SERB, India, through grant numbers CRG/2021/005320, ECR/2016/001004, and the use of high-performance computing facilities at IISER Bhopal.
\end{acknowledgments}
% ====================

% ========== BIBLIOGRAPHY ==========
%\bibliography{/Users/nirmal/Documents/bibliography/library,library}

\end{document}